\documentstyle[preprint,tighten,aps]{revtex}

\begin{document}
\draft
\preprint{\vbox{\noindent
          \null\hfill  INFNCA-TH9816}}
\title{Thermal distributions in stellar plasmas, \\
nuclear reactions and solar neutrinos}
\author{M.~Coraddu$^{1,2,}$\cite{email1},
        G.~Kaniadakis$^{3,}$\cite{email2},
        A.~Lavagno$^{3,4,}$\cite{email3},
        M.~Lissia$^{2,1,}$\cite{email4},\\
        G.~Mezzorani$^{1,2,}$\cite{email5},
    and P.~Quarati$^{3,2,}$\cite{email6}}
\address{
$^{1}$Dipartimento di Fisica, Universit\`a di Cagliari,
                I-09042 Monserrato, Italy\\
$^{2}$Istituto Nazionale di Fisica Nucleare, Sezione di Cagliari,
                I-09042 Monserrato, Italy\\
$^{3}$Dipartimento di Fisica and INFM, Politecnico di Torino,
      I-10129 Torino, Italy \\
$^{4}$Istituto Nazionale di Fisica Nucleare, Sezione di Torino,
        I-10125 Torino, Italy
        }
\date{November 23, 1998}
\maketitle

\begin{abstract}
The physics of nuclear reactions in stellar plasma is reviewed with
special emphasis on the importance of the velocity distribution of
ions. Then the properties (density and temperature) of the weak-coupled
solar plasma are analysed, showing that the ion velocities should deviate
from the Maxwellian distribution and could be better described
by a weakly-nonexstensive ($|q-1| < 0.02$) Tsallis' distribution.
We discuss concrete physical frameworks for calculating this deviation:
the introduction of higher-order corrections to the diffusion and
friction coefficients in the Fokker-Plank equation,
the influence of the electric-microfield stochastic distribution on
the particle dynamics, a velocity correlation function with long-time
memory arising from the coupling of the collective and individual
degrees of freedom.
Finally, we study the effects of such deviations on stellar nuclear rates,
on the solar neutrino fluxes, and on the $pp$ neutrino energy spectrum, and 
analyse the consequences for the solar neutrino problem. 
\end{abstract}

\section{Introduction}
\label{secintro}
Thermonuclear reactions are of enormous importance for the physics of
stars and of our Sun. The general formalism that describes these
reactions exists since a long time and there is a wide consensus about
our good understanding of the relevant
physics~\cite{clayton68,rolfs88,bahcall89}.

However, quantitative calculations of specific reaction rates need
experimental inputs and theoretical assumptions. {\em Cross sections must
be known}. In the best cases, they can be directly measured at the relevant
energies; other cross sections can be only measured at higher energies and
need to be extrapolated; in some situations only theoretical
predictions exist (for a recent review of the cross sections relevant to
the Sun see Ref.~\cite{INTW98}). 
In addition, rates depend on weighted thermal averages of the
cross sections; therefore, knowledge about {\em the thermal distribution is
also needed}.

Because many nuclear reactions in the stellar burning core proceed by way of
quantum penetration of a high Coulomb barrier, their cross sections grow
exponentially with energy. Therefore, thermal averages do not
probe the average energy of the distribution ($kT$), but its
high-energy tail. Consequently, {\em rates are sensitive to a relatively
small part of the distribution}.

The ion velocity distribution in stellar calculations is always
assumed Maxwellian~\footnote{We consider situations where quantum
effects are negligible, {\em e.g.}, in the solar plasma small quantum
corrections to the statistics exist for electrons but not for ions. More
generally, one could also analyse plasma corrections to the
standard Fermi or Bose distributions.}. In fact,
it is almost universally accepted as a fact that physical conditions
in the solar interior (density and temperature) lead to an equilibrium
velocity distribution that is Maxwellian.
In this article we review the assumptions underlying
this general statement, discuss why they are only approximately true,
and argue that the consequent {\em small corrections to the distribution
have significant effects on the rates}.

The dynamics of the solar plasma is not trivial, since at such
densities and temperatures there is no clear scale separation
between collective and individual degrees of freedom~\cite{kania97}. 
The presence of
more than just one energy scale ($kT$) in the relevant range of energies
results in deviations from the pure exponential behavior, $\exp{(-E/kT)}$, 
which is determined only by $kT$.
We have tried several new approaches to this strongly interacting
many-body system for which a realistic microscopic calculation does
not exist yet.

In one approach~\cite{kania93,kania97b}
we exploit the knowledge that the distribution in the
solar interior cannot be too much different from the Maxwellian one
and add small corrections (higher-order terms in a derivative expansion)
to the coefficients of the standard Fokker-Plank equation. 
Tsallis'~\cite{tsallis88,curado91} 
and Druyvenstein-like distributions are immediately generated.

A second approach~\cite{gervino98}
focuses on the electric microfields that have been
shown to exists in plasmas and tries to link their distribution
and the connected effective cross sections with deviations from
the Maxwellian distribution. Different classes of microfields
distributions and effective cross sections yield several
non-Maxwellian distributions among which we again find Tsallis and
Druyvenstein-like distributions.

The third approach has been just started and aims to connect the distribution
of collective variables~\cite{valuev98}
to memory effects and long-time correlations
between velocities. There should exist solutions compatible with
the Tsallis' distribution and/or other non-Maxwellian distributions.

These three approaches are not exhaustive and not necessary alternative.
Nevertheless, it is suggestive that all of them point in the same direction:
{\em the Maxwell-Boltzmann distribution of velocity should have small but
nonnegligible corrections in the solar plasma and the Tsallis' distribution
could provide a better description}.

Nowadays solar modeling seems to have reached a satisfactory
stage~\cite{clayton68,bahcall89,bahcall95a}.
The inclusion in the latest models of higher-order effects, such as the
diffusion of heavy elements, brings the theoretical predictions in good
agreement even with the detailed helioseismological
data~\cite{bahcall97b,fiorentini97}. However, there still exist a
discrepancy between the solar neutrino experiments and the predicted
neutrino fluxes~\cite{bahcall95b}.
In this context, there has been a considerable amount of work devoted
to answering questions such as: how large are the uncertainties of the
solar model input parameters? Has something been left out of standard
solar models? How does this affects predictions for the fluxes and the
status of solar neutrino
problem (SNP)~\cite{Krauss93,Fogli94,hata94b,fogli95,bahcall96,castell97}?

Therefore, we find extremely important to assess the consequences of the
possible small deviations from the standard statistics on the solar model and,
in particular, on the neutrino fluxes. At least two effects should be
considerd: changes of the total rates caused by the different thermal
average and modifications of the shape of the energy spectra. The effect
on the rates was already considered by
Clayton~\cite{clayton74,clayton75} two decades ago. New
experimental data, better solar models and a considerable better
understanding of nonstandard distributions and of how they can arise
in solar plasmas convinced us of the necessity of reconsider this 
earlier suggestion. Even if experimental detection of small modifications
of the neutrino spectra might appear still far from being possible,
it is useful to have a clear assessment of what the signal would be.

Our paper is organized as follows. In Sec.~\ref{secrate} we review the
physics of subbarrier thermonuclear reactions and how the velocity
distribution influences reaction rates. Section~\ref{secplasma}, which
is dedicated to the solar plasma, contains the central part of our work:
three approaches to velocity distribution calculation and the resulting
nonstandard distributions. In Sec.~\ref{secmodirate} we calculate the
reaction rates with the modified distributions, while we obtain
the neutrino spectrum from the $pp$ reaction in the presence of
Tsallis' statistics and compare it to the standard one in
Sec.~\ref{secspectrum}. The solar neutrino problem is analysed in the
light of our results on nonstandard velocity distributions in
Sec.~\ref{secSNP}, and Sec.~\ref{secconclu} is reserved to our conclusions.

\section{Thermonuclear reaction rates}
\label{secrate}
Reliable calculations of nuclear reaction rates in stellar interiors is
fundamental for a quantitative understanding of the structure and
evolution of stars. In fact, while the overall stellar structure is rather
robust, changes of some of the rates by few percent can produce detectable
discrepancies, when precise measurements are possible, {\em e.g.}, in the
case of the solar photon and neutrino luminosity, and mechanical
eigenfrequencies~\cite{degli98}.

In this Section, we review the basic physics underlying thermonuclear 
reactions and the main ingredients of their calculation~\cite{clayton68}.

\subsection{Introduction}
Let us consider a gas with $n_{1}$ particles of type 1 and $n_{2}$
particles of type 2 per cubic centimeter and relative velocity $v$;
the reaction rates $r$ (the number of reactions per unit volume and unit time)
is given by
\begin{equation}
r= (1+\delta_{12})^{-1}\, n_1 n_1 \langle v \sigma \rangle \ \ ,
\label{rates}
\end{equation}
where $\sigma=\sigma (v)$ is the nuclear cross section of the reaction.
The reaction rate per particle pair is defined as
\begin{equation}
\langle v \sigma \rangle = \int_0^\infty \! f(v)\,  \sigma v \, dv \, ,
\label{sigmav}
\end{equation}
where the particles distribution function $f(v)$ is a local function
of the temperature.

Therefore, the reaction rate per particle pair $\langle v \sigma \rangle$
is determined by the specific cross section and by the velocity distribution
function of the incoming particles. In general, cross sections do not have
very strong dependence on the energy, when no energy barrier is present and
away from resonances. Therefore, most of the contribution to
$\langle v \sigma \rangle$ comes from particles with energy of the order of
$k T$, and the dependence on the specific form of $f(v)$ is weak. We
shall see that the situation is very different in the presence of the Coulomb
barrier.
\subsection{Subbarrier reactions}
Most of the nuclear reactions that power the stars are between charged
particles. These particle must penetrate a Coulomb barrier that
is very large in units of $k T$. 
The penetration probability is proportional to the Gamow factor
$\exp{[-2\pi\eta(E)]}$, where $\eta(E)=Z_1 Z_2 \alpha \sqrt{\mu c^2/2E}$
is the Sommerfeld parameter, $\alpha$ is the fine structure constant and
$\mu$ is the reduced mass; this factor can be also written as
$\exp{[-\sqrt{E_G/E}]}$, defining the Gamow energy 
$E_G= 2\mu c^2 (Z_1 Z_2 \alpha \pi)^2 $. This exponentially small
probability makes the cross section grow extremely fast with the energy;
therefore, one usually defines the astrophysical $S$ factor, whose energy
dependence is weaker
\begin{equation}
\label{sigmas}
\sigma (E)=\frac{S(E)}{E} e^{-2 \pi\eta(E)} =
           \frac{S(E)}{E} e^{-\sqrt{E_G/E}}  \ \ .
\end{equation}

This reaction mechanism has at least two main consequences for the
study of stellar structure.

(1) Since the bulk of particles in the stellar plasma have thermal energies
of the order of $kT$, which is far below the Coulomb barrier, only a small
number of particles in the high-energy tail of the distribution has a
chance of reacting: 
this high-energy tail plays a crucial r\^{o}le for the reaction rates,
which becomes very sensitive to small changes of the distribution.

(2) The cross sections of some of the reactions, {\em e.g.}, weak reactions
with Coulomb barrier such as $p+p\to d + e^{+} +\nu$, are extremely small at
the low energies relevant to stellar physics (of the order of tens of keV for
reactions in the Sun). Therefore, some of them have only been calculated
theoretically, others have only been measured at higher energies and then
extrapolated to the lower thermal energies of the stellar interiors.

Both these two facts lead to considerable uncertainties in the reaction
rates. The second point has been extensively discussed, and it is common
opinion that the present standard theory of stars, and in particular of
the Sun~\cite{castell97,INTW98}, has already taken properly into account
the uncertainties in the cross sections.
On the contrary, the first
point, the possibility that small changes of the particle energy 
distribution could strongly affect reaction rates, has received little
attention~\cite{clayton74,haubo95,kania96,kania97}, 
or dismissed on the ground
that the energy distribution in stars is allegedly well-know~\cite{bahcall89}. 
In this paper, we shall
discuss mainly this point, the energy distribution in stellar plasmas and
its implication for nuclear rates.

\subsection{Maxwell distribution and the Gamow peak}

Normal stellar matter, such as the one in the Sun, is nondegenerate,
{\em i.e.}, quantum effects are small (in fact, they are small for electrons
and completely negligible for ions), it is nonrelativistic, and it is in
good thermodynamical equilibrium. On this ground, the particle velocity
distribution is almost universally taken to be 
a Maxwell-Boltzmann distribution, without much questioning.

However, derivations of the ubiquitous Maxwell-Boltzmann distribution are
based on several assumptions~\cite{kania97}.
In a kinetical approach, one assumes (1)
that the collision time be much smaller than the mean time between collisions,
(2) that the interaction be sufficiently local, (3) that the velocities of
two particles at the same point are not correlated (Boltzmann's
Stosszahlansatz), and (4) that energy is locally conserved when using only the
degrees of freedom of the colliding particles (no significant amount of
energy is transferred to collective variables and fields). In the 
equilibrium statistical mechanics approach, one uses the assumption
that the velocity probabilities of different particles are independent,
corresponding to (3), and that the
total energy of the system could be expressed as the sum of a term quadratic
in the momentum of the particle and independent of the other variables,
and a term independent of momentum, but if (1) and (2) are not valid
the resulting effective two-body
interaction is not local and depends on the momentum and energy of
the particles.
Finally, even when the one-particle distribution is Maxwellian,
additional assumptions about correlations between particles are necessary
to deduce that the relative-velocity distribution, which is the
relevant quantity for rate calculations, is also Maxwellian.

At least in one limit the MB distribution can be rigorously derived:
systems that are
dilute in the appropriate variables, whose residual interaction is
small compared to the one-body energies. In spite of the fact
that the effects of the residual interaction cannot be neglected,
as a good first approximation the solar interior can be studied in this dilute
limit; therefore, it is reasonable to suppose that the velocity distribution
in the Sun is not too far from the Maxwellian one.
This fact is consistent with the many successes of the standard treatment
and suggests that we start discussing the standard results and improve
them later.

In the ordinary treatment, the single particle energy distribution
for protons and other ions is taken as 
\begin{equation}
\label{MBdis}
f_{MBD}(E) =\frac{2} {\sqrt{\pi}} \,
\, \frac{\sqrt{E}}{(kT)^{3/2}} \, e^{-E/kT} \, .
\end{equation}
When this distribution and a cross section of the form of Eq.~(\ref{sigmas})
are inserted in Eq.~(\ref{sigmav}), the resulting integrand goes to
zero both at large energies, because of the exponentially small number of
particles, see Eq.~(\ref{MBdis}), and also at small energies, because of
the exponentially small probability of barrier penetration, see
Eq.~(\ref{sigmas}). In fact, the integrand has a maximum at the
temperature-dependent energy
\begin{equation}
\label{mosteffen}
E_0 = \left( \frac{E_G (k T)^2}{4} \right)^{1/3} = \tau \, k T\, ,
\end{equation}
which it is called the most effective energy~\cite{clayton68}, since
most of the particles that react have energies close to $E_0$; we have
defined the adimensional parameter
$\tau = E_0/kT = [E_G/(4 k T)]^{1/3}$.

Figure~\ref{figgamow} gives a pictorial demonstration of how the Gamow peak
originates. The exponentially decreasing function
(solid curve labeled ``Maxwell'') is the Maxwellian factor
$\exp{(-E/kT)}$ multiplied times 400 to emphasize the tail of the
distribution. The rapidly growing function (dotted curve labeled
``Penetration factor'') is the factor $\exp{-\sqrt{E_G/E}}$ with the choice
of $E_G =  4000 kT$, which corresponds to a most effective energy
$E_0 = (E_G (kT)^2 / 4)^{1/3} = 10 kT$; it has been arbitrarily normalized such
that it is equal to 400 at $E=20kT$. The product of the two functions
yields the Gamow peak (solid curve labeled Gamow), which has been
normalized to one at its maximum ($E=E_0 = 10 kT$).
It is important to notice that the area under the Maxwellian curve for
energies larger than about $6 kT$ (where one finds most of the contribution
to the peak) is less then 0.3\% of the total area.

In this framework, one performs the integral in Eq.~(\ref{sigmav}) using
the saddle-point asymptotic expansion around the maximum $E_0$ of the Gamow
peak and finds that the reaction rate per particle pair is, apart for small
calculable corrections,
\begin{equation}
\label{asyrate0}
\langle v \sigma \rangle_{12} = 
\frac{8}{\sqrt{3} \mu_{12} c \pi \alpha Z_1 Z_2}
S(E_0) \tau_{12}^2 e^{-3\tau_{12}} \, .
\end{equation}

Table~\ref{table1} reports the values of $E_0$ in units of $k T$
($\tau$) for several reactions. We notice that $E_0$ can be much larger than
$k T$ and, therefore, only a very small number of particles in the far
tail of the distribution contributes to the rate, {\em e.g.}, for the 
${}^3$He + ${}^3$He reaction the most effective energy is about 
$17 k T$: there are less than about 40 particles out of a million that
have energies so large or larger. It is not sufficient
anymore to know that the Maxwellian distribution is good approximation,
we must be sure that there are no corrections to a very high accuracy.
\section{Velocity distributions in non-ideal plasmas}
\label{secplasma}
As discussed in Sec.~\ref{secrate}, several subtle points must be assumed
in the derivation of the often-taken-for-granted Maxwellian
velocity distribution. Likewise, one often assumes that the solar core
could be treated as an ideal (Debye) plasma. However, there are physical
conditions and/or specific applications that needs higher accuracy for
which becomes necessary to take into account modifications of the
standard plasma theory.

In this Section we discuss how the physics of non-ideal plasmas, and in 
particular of the solar interior, can result in equilibrium velocity
distributions that deviate from the ``standard'' one. In addition, we
present specific physical frameworks where non-Maxwellian distributions
arise and microscopic mechanisms that allow a reliable estimation of the
size of deviations.

\subsection{Ideal and non-ideal plasma}

In literature, a plasma is characterized by the value of the plasma parameter 
$\Gamma$ 

\begin{equation}
\Gamma=\frac{(Z e)^2}{a \, kT}  \ \ ,
\end{equation}
where $a = n^{-1/3}$ is of the order of the interparticle average distance
($n$ is the average density). The plasma parameter is a measure of the ratio
of the mean (Coulomb) potential energy and the mean kinetic (thermal) energy.

Depending on the value of the plasma parameter, we can distinguish 
three regimes that are characterized by different effective interactions
and require different theoretical approaches.

\begin{itemize}
\item 
$\Gamma \ll 1$. The plasma is described by the Debye-H\"uckel
mean-field theory as a dilute weakly interacting gas.
The screening Debye length 
\begin{equation}
R_D=\sqrt{ \frac{kT}{4\pi e^2 \sum_i Z^2_i n_i} } \ \ ,
\end{equation}
is much greater than the average interparticle distance $a$, hence there
is a  large number of particles in the Debye sphere
($N_D\equiv(4\pi/3)R_D^3$).
Collective degrees of freedom are present (plasma
waves), but they are weakly coupled to the individual degrees of freedom
(ions and electrons) and, therefore, do not affect their distribution.
Binary collisions through screened forces produce
the standard velocity distribution.

\item 
$\Gamma \approx 0.1$. The mean Coulomb energy potential is not much
smaller of the thermal kinetic energy and the screening length 
$R_D\approx a$. It is not possible to clearly separate individual and
collective degrees of freedom. The presence of at least two different
scales of energies of the same rough size produces deviations from the
standard statistics which describe the system in terms of a single
scale, $kT$.

\item 
$\Gamma > 1$. This is a high-density/low-temperature plasma where the
Coulomb interaction and quantum effects start to dominate and determine the
structure of the system.
\end{itemize}

\subsection{The solar interior}

In the solar interior the plasma parameter is
$\Gamma_\odot\approx 0.1$; therefore, the solar core is a weakly non-ideal
plasma where the Debye-H\"uckel conditions are only
approximately verified. Similar behaviors are expected in other
astrophysical systems characterized by plasma parameters in the range
$0.1\le \Gamma\le 1$; examples are brown dwarfs, the Jupiter core
and stellar atmospheres.

Studies of systems with this intermediate values of $\Gamma$ are the most
difficult;
it is possible to use/combine several different approaches none of which,
however, is completely justified.

The reaction time necessary to build up screening after a hard collision can
be estimated from the inverse solar plasma frequency 
$t_{pl}=\omega_{pl}^{-1} = 
\sqrt{m/ (4\pi n e^2 )} \approx 10^{-17}$~sec, 
and it is comparable to the
collision time $t_{coll} = \langle\sigma v n \rangle^{-1} \approx 10^{-17}$ 
~sec. Therefore, several collisions are likely necessary before the particle
looses memory of the initial state and the scattering
process can not be considered Markovian. In addition, screening
starts to become dynamical: the time necessary to build up again the
screening after hard collisions is not negligible any more.

\subsection{Three roads towards nonstandard distributions}
For concreteness we consider three possible approaches to
weakly-noninteracting plasmas that yield
nonstandard distributions. These approaches are not alternative, but
they could perhaps be seen as different and partial descriptions of the same
complicated physical problem. The main purpose of their presentation is
to give concrete examples of how nonstandard statistics arises. However,
they are not by any mean the last word on the topic: we must still
develop their full potentiality and alternative approaches must also
be pursued. 

\subsubsection{Fokker-Plank}
In the Fokker-Plank context it is possible to introduce
corrections (i) to the lowest order friction coefficient $J(v)$, and (ii)
to the lowest order diffusion coefficient 
$D(v)$~\cite{kania93,kania97b,kania97}.
Similar corrections have already been shown to exist in hydrodynamic
systems.

 {\em We assume that the system is not too far from the
  standard regime that leads to the MB distribution, so that an expansion
starting from the usual formalism makes sense}. The Fokker-Planck equation,
given in the Landau form, is
\begin{equation}
\frac{\partial}{\partial t}f(t,v) = \frac{\partial}{\partial v}
\left(
J(v)f(t,v) + \frac{\partial}{\partial v} D(v) f(t,v)
\right)\, ,
\end{equation}
where $f(t,v)$ is the distribution probability of particles with velocity
$v$ at time $t$ and $J(v)$ and $D(v)$ are
the dynamical friction and diffusion coefficients.
The stationary distributions are the asymptotic solutions of the above equation.
To lowest order $J(v)= v/\tau$ and $D(v)=\epsilon/\tau$, where the constant
$\tau>0$ has dimension of time ($m / \tau$ is the friction constant) and
$\sqrt{\epsilon}$ has dimension of a velocity
($\epsilon = kT/m$ for Brownian motion).
At equilibrium one obtains the well-known Maxwellian distribution
\begin{equation}
f(v)\equiv \lim_{t\to\infty} f(t,v) \sim
\exp\left\{-\frac{v^2}{2\epsilon}\right\}
= \exp\left\{-\frac{m v^2}{2 kT}\right\} \, .
\end{equation}

We can generalize the standard Brownian kinetics considering the expressions
of the quantities $J(v)$ and $D(v)$ to the next order in the velocity
variable: $J(v)= v/\tau\, (1+\beta_1 v^2/\epsilon) $ and
$D(v)=\epsilon/\tau \, (1 + \gamma_1 v^2/\epsilon)$; these higher derivative
terms can be interpreted as signals of nonlocality in the Fokker-Planck
equation.

If $\beta_1= 0$ and $\gamma_1\neq0$ we find the Tsallis' distribution
\begin{equation}
\label{tsdis}
f(v) = \left[ 1 + (q-1)\frac{m v^2}{2 kT}
      \right]^{1/(1-q)}
       \Theta\left[1 + (q-1)\frac{m v^2}{2 kT}\right]\, ,
\end{equation}
where $q-1=2\gamma_1/(2\gamma_1+1)$, $\Theta$ is the
Heaviside step-function, and $kT/m \equiv \epsilon (2-q)$.
When the characteristic parameter $q$
is smaller than 1 ($-1/2<\gamma_1<0$), this distribution has a upper cut-off:
$m v^2/2 \leq kT/(1-q)$ (the tail is depleted). The distribution correctly
reduces to the exponential Maxwell-Boltzmann distribution in the limit
$q\to 1$ ($\gamma_1\to 0$).
When the parameter $q$ is greater than 1 ($\gamma_1>0$), there is no cut-off
and the (power-law) decay is slower than exponential
(the tail is enhanced).

If $\beta_1\neq 0$ and $\gamma_1=0$, we find a
Druyvenstein-like distribution:
\begin{equation}
f(v)\sim
\exp\left\{  - \frac{v^2}{2\epsilon}
            - \beta_1 \left( \frac{v^2}{2\epsilon} \right)^2
   \right\} \, ,
\end{equation}
which has also the functional form suggested by
Clayton~\cite{clayton75}
to parameterize a small deviation (depletion) from the Maxwellian statistics.

\subsubsection{Random fields}
Each particle is affected by the total electric field distribution due to the
other charges in the plasma. If external fields and large scale internal fields
due to collective modes are subtracted from the total electric field, the
single particle see the remaining field as a relative small random component.
The density of these random electric microfields has been
studied~\cite{igle83,roma98} and it
is often expressed in terms of the dimensionless parameter $F$
as $\langle {\cal E}^2 \rangle = (F \, e/a^2)^2$, where $4\pi a^3/3 = 1/n$;
the distribution of $F$ in plasmas depends
on the value of $\Gamma$~\cite{roma98}.
These microfields have in general long time correlations, and can
generate anomalous diffusion.

The total (micro)field can be decomposed into three main components.\\
(i) A slow-varying (relative to the collision time) component due to
collective plasma oscillations, which the particle sees as an almost
constant external mean field ${\cal E}$ over several collisions.\\
(ii) A fast random component due to particles within a few Debye radii,
whose effect can be described by an elastic diffusive cross section
$\sigma_1\sim v^{-1}$. When only this cross section is present, the
distribution remains Maxwellian even in presence of the slow mean field
${\cal E}$.\\
(iii) A short-range two-body strong Coulomb effective interaction, that
can be described by the ion sphere model~\cite{ichi86}. The strict enforcement
of this model as implemented by Ichimaru~\cite{ichi86} yields the elastic
cross section $\sigma_0=2\pi\alpha^2a^2$, where $a$ is the interparticle
distance, $\alpha$ an adimensional parameter whose order of magnitude can
be inferred from the parameter $F$ that characterizes
the microfields, $F\approx \alpha^{-2}$. 
Since $F^2 \sim 3/\Gamma\approx 40$, for $\Gamma = 0.07$, one can estimate
$0.4<\alpha<1$. In the present model, it is this component of the electric
field that turns out to be mainly responsible of the correction factor
$\exp{[-\hat{\delta} (E/kT)^2]}$.

In this framework, the stationary solution of the kinetic equation valid for
small deviations from the MB distribution can be shown to be the Druyvenstein 
distribution:
\begin{equation}
f(E) \sim \exp{\left[ - \hat{\varphi} \frac{E}{kT}
              - \hat{\delta}  \left(\frac{E}{k T}\right)^2 
               \right]}  \, ,
\label{fmicro}
\end{equation}
where 
\begin{equation}
\hat{\varphi}=\frac{\varphi}{1+\varphi} \ \ \ \ \ \varphi=\frac{9}{2} \kappa
\left(\frac{n k T}{Z e {\cal E}}\right)^2 \langle\sigma_1^2\rangle \, , 
\nonumber
\end{equation}
\begin{equation}
\hat{\delta}=\left (\frac{3 \langle\sigma_1^2\rangle}{\sigma_0^2} + 
\frac{1}{\delta}\right )^{-1} \ \ \ \ \ \delta=\varphi
\frac{\sigma_0^2}{3 \langle\sigma_1^2\rangle} \, , \nonumber 
\end{equation}
$\kappa=2 \, m_a m_b/(m_a+m_b)^2$ is the elastic energy-transfer
coefficient between two particles $a$ and $b$, and
$\delta$ is proportional to the square of the ratio of the energy densities
of the electric field and of the thermal motion.

In the small correction limit, relevant to the solar interior,
$\hat{\varphi}=1$ and the $\hat{\delta}$ parameter is: 
\begin{equation}
\label{deltamodel}
\vert\hat{\delta}\vert \approx 
\frac{\sigma_0^2}{3 \langle\sigma_1^2\rangle}=12 \, \alpha^4 
\, \Gamma^2 \ll 1 \, .
\end{equation}

From Eq.~(\ref{fmicro}), we see that the presence of electric microfields
implies 
a deviation from the Maxwell-Boltzmann distribution and that the entity of this 
deviation depends on the value of the plasma parameter. As already mentioned, 
in the solar core $\Gamma_\odot\approx 0.1$, hence, the order of magnitude of 
the deviation parameter $\hat{\delta}$ is about $0.01$ and in this
intermediate region long-range and memory effects take place. 

\subsubsection{Memory effects and collective variables}

As already discussed, in the solar core collective effects have
time scales comparable to the average time between collisions
({\em e.g.}, compare the inverse plasma frequency with the average 
collision time $\omega_{pl}^{-1}\approx t_{coll}$) and it is not possible
a description that separates the collective and the individual degrees
of freedom.

A rigorous microscopic approach would imply the resolution of the dynamical 
equations of motion with a Hamiltonian that explicitly contains collective
and individual degrees of freedom and their mutual interactions.
Our scope is more modest: we want to give a concrete example where memory
effects (long-time correlations) are important.

The authors of Ref.~\cite{bohm64,valuev98} introduce
the collective variables
\begin{equation}
\rho_k (t)=\sum_j^N \exp(-ik x_j(t)) \ \ ,
\end{equation}
and describe these collective variables as harmonic oscillator variables
weakly coupled to the individual degrees of freedom, which, therefore,
act as a thermostat for the collective variables.

Generalizing this approach, we write down a generalized Langevin
equation for the collective variables
\begin{equation}
\ddot{\rho}_k=-\omega_k^2 \rho_k+
\int_0^t K(t-\tau) \dot{\rho}_k (\tau) d\tau+F(t) \ \ ,
\end{equation}
where $K(t-\tau)$ is the memory friction kernel, which takes into account
the long-time tail of the correlations in the solar interior, and $F(t)$
is the stochastic fluctuating thermal force due to the interactions
with the individual degrees of freedom.
In weakly non-ideal plasmas, the longitudinal waves have
the following dispersion law 
\begin{equation}
\omega^2_k=\omega_{pl}^2 (1+3 R_D^2 k^2) \ \ .
\end{equation}

The two functions $K(t-\tau)$ and $F(t)$ are not independent but
related by the second fluctuation-dissipation theorem
\begin{equation}
\langle F(t_1) F(t_2)\rangle=\frac{2 kT}{m}\, K(t_1-t_2) \ \ .
\label{sfd}
\end{equation}
In the limit of negligible memory effects, the memory kernel
becomes a $\delta$-function, that is, $K(t)=\gamma\,\delta (t)$, 
where $\gamma$ is the effective friction coefficient, which can be written
in terms of the Landau ($\gamma_L$) and 
collisional ($\gamma_c$) damping~\cite{valuev98}. 

The explicit form of the memory kernel depends on the specific system
and, in principle, should be deduced from a microscopic 
calculation. At the moment, we have been able to verify that, in the
framework of Tsallis  nonextensive statistics, the memory kernel can be
written, for $q\approx 1$, as
\begin{equation}
K(t)=\gamma\,\delta (t)+(1-q) \gamma^2 
\left (1+2\gamma t+\frac{\gamma^2  \,t^2}{2}\right ) \ \ .
\end{equation}

The above expression is consistent with the prescription of the nonextensive 
statistics and implies anomalous subdiffusion for $q<1$ and superdiffusion for 
$q>1$. A more complete analysis of this possibility is still under way; it
is also important to study the effects of such a memory kernel
for the collective modes on the individual degrees of freedom.

At the moment, we can speculate that such correlations among the
collective modes could
lead to a long-time asymptotical behavior of the velocity-correlation
of the ions of the kind $\langle v(0)v(t)\rangle\sim t^{-(1+\gamma)}$.
If $\gamma\geq 1$, {\em i.e.},
the correlation decays sufficiently fast, the diffusion process is
no qualitative different from the delta-function case:
$\langle x^2(t)\rangle\sim t$. In this case, one can
show~\cite{muralidhar90,wang92} that if $0<\gamma<1$ and
$0<\int\langle v(0)v(t)\rangle<\infty$, the standard
distribution of velocity is still valid. However, if $0<\gamma<1$ 
and $\int\langle v(0)v(t)\rangle=0$ (or very small),
or if $-1<\gamma\leq 0$, the diffusion is anomalous 
$\langle x^2(t)\rangle\sim t^{1+\gamma}$ ($\sim t\log t$, if $\gamma=0$).
Indeed, Tsallis~\cite{tsallis95c} shows that the generalized entropy
quite naturally generate both anomalous diffusion
($\langle x^2(t)\rangle\sim t^{1+\gamma}$) and the non-Maxwellian
probability distribution for the velocities of Eq.~(\ref{tsdis}).

\subsection{Summary}
In this Section we have presented theoretical arguments that the
velocity distribution in the solar core should deviate from the
standard one and, in particular, that it could follow Tsallis'
distribution~\cite{tsallis88,curado91} 
shown in Eq.~(\ref{tsdis}). We have also estimated
the deviation from the standard statistics from the known microfield
distribution. If we use the
parameterization suggested by Clayton~\cite{clayton74,clayton75}
(Druyvenstein distribution)
\begin{equation}
f(E) \sim (kT)^{-3/2} \, 
       e^{-E/kT - \hat{\delta} (E/kT)^2 } \, ,
\end{equation}
$\hat{\delta}$ should be of the order of 0.01. Since in the limit $q\to 1$
the Tsallis distribution can be asymptotically described as a Druyvenstein
with $\hat{\delta} =(1-q)/2$, we have also estimated that $q$ should be
a few percent different from 1.

In the next section, we shall demonstrate how even such small deviations 
from the Maxwell-Boltzmann distribution can be very important for
solar physics.

\section{Reaction rates and modified thermal distribution}
\label{secmodirate}
Both from general considerations about the successes of the standard
approach and from the estimates and calculations that we have shown, it
should be clear that the deviations from the Maxwellian distribution
in the Sun are small. Therefore, it is completely general to make an
asymptotic expansion of the following kind:
\begin{equation}
f(E) \sim (kT)^{-3/2} \, 
       e^{-E/kT - \delta (E/kT)^2 } \, ,
\end{equation}
where we have disregarded terms with powers higher than $(E/kT)^2$ in the
exponent and all other power corrections outside the exponential, apart
the ones (not shown) needed to correctly normalize the distribution.
This result becomes the more accurate the more $\delta$ is small.
Note that the exponential can not be expanded.

This same kind of parameterization was already considered by
Clayton~\cite{clayton75}. 
One can easily convince oneself that the distributions
previously discussed can be put in this form in the limit of small deviations.
We have seen that the the Tsallis' distribution plays a special r\^{o}le: it
can be also approximated to first order in $(1-q)$ by Clayton's form with
$\delta=(1-q)/2$ and a renormalized temperature $T'= T + T(1-q)$. 
In Sec.~II we performed a saddle point expansion of the integral over
the velocity distribution with the introduction of the  most effective
energy, $E_0$, and approximating the integrand around the maximum $E_0$ with
a Gaussian function (Gamow peak). The same kind of asymptotic expansion can
be repeated for this modified distribution yielding the following
analytical expression for the change of the rate:
\begin{equation}
\label{asyratedel}
\frac{\langle v\sigma_i\rangle_{\delta}}{\langle v\sigma_i\rangle_{0}}=
e^{-\delta (E_0^{(i)}/kT)^2}\equiv e^{-\delta\gamma_i}    \, ,
\label{edg}
\end{equation}
where $E_0$ is the most effective energy of Eq.~(\ref{mosteffen})
\begin{equation}
\frac{E_0}{kT}\approx 5.64 \left(Z_1^2Z_2^2\frac{A_1A_2}{A_1+A_2}
            \frac{T_c}{T}\right)^{1/3} \, ,
\end{equation}
which depends on the reaction $i$, through the charges $Z$ and weights $A$ of
the ions, and on the relevant average temperature $T$ ($T_c=1.36$~keV is
the temperature at the center of the Sun), and where
$\langle v\sigma_i\rangle_{0}$ is the expression in Eq.~(\ref{asyrate0}).

We shall use the expression in Eq.~(\ref{asyratedel}) to discuss the effects
of modified
statistics on the solar thermonuclear rates: it has a sufficient numerical
accuracy for this situation and it is physically more transparent.
However, we perform the integral over the distribution numerically
in physical situations where stronger deviations from the Maxwellian
distribution are expected, and when we want to check whether the asymptotic
expansion is sufficiently accurate.

The dramatic effect of small deviations from the Maxwellian distribution
on the rate can be appreciated by looking at Fig.~\ref{figgamow}. The
steep decreasing dashed curve shows the high-energy tail of the Tsallis'
distribution $ [1 + (q - 1)E/kT]^{1/(1 - q)}$ with $q=0.98$
compared with the corresponding Maxwellian distribution (solid curve).
The tails of the two distributions shown in Fig.~\ref{figgamow}
contain about 0.30\% (Maxwell) and 0.26\% (Tsallis) of their total area and 
have been multiplied times a factor 400 to make their small difference
more visible. The product of each of these two very similar distributions
times the same penetration factor (dotted curve) yields two very
different Gamow peaks. The peak corresponding to the Tsallis distribution
(dashed curve) is not only shifted at lower energy by more than one unit
of $kT$ compared to the Maxwellian peak (solid curve), but it is also
much lower: {\em a tiny change of the distribution tail 
($4\times 10^{-4}$ less probability in the tail) makes the probability
in the peak smaller by a factor of about 0.3!}

\section{Thermal effects on the neutrino energy spectrum}
\label{secspectrum}
Because of their small interaction cross section, neutrinos that
reach the earth have not interacted with the solar matter and carry
direct information from the solar core. One might hope that their
energy spectrum could tell us something about the particle velocity
distribution in the Sun. 

The neutrino energy spectrum for a given
reaction depends on the total energy available in the center of mass
to the final products, total energy that is the sum of the thermal
energy of the
incoming particles and of the energy released by the reaction, and on the
velocity of the center of mass that produces Doppler broadening. As a
specific example, we shall use the $pp$ fusion reaction, since its neutrino
spectrum is easily calculable from phase space and since it releases a
relatively low excess energy leaving more space for thermal effects.

The reaction
\begin{equation}
p + p \to d + e^{+} + \nu_{e}
\end{equation}
releases an excess energy of $Q=2 m_{p} - m_{d} - m_{e} = 2\times 938.272
- 1875.613 - 0.511 = 0.420$~MeV, which is shared among the
kinetic energies of the outgoing particles.
Note the the total energy released by the reaction is
$1.442$~MeV, since the positron annihilation gives additional $1.022$~MeV;
this energy, however, does not contribute to the kinetic energy of
the neutrino (this should be contrasted to the more rare reaction
$p + p + e \to d + \nu_{e}$).

Since there are three particles in the final state, the neutrino energy in
the center of mass has a known probability distribution. For simplicity,
we report only the result that disregards the recoil energy (a relative
correction of the order of $2\times 10^{-4}$)
\begin{eqnarray}
\label{spectrpp}
P_{K}(E_{\nu}) & \sim & E_{\nu}^2 (K+Q+m_{e}-E_{\nu}) 
          \sqrt{(K+Q -E_{\nu})(K+Q+2m_{e}-E_{\nu})} \times F(E_{\nu}) 
          \nonumber \\
         &=& E_{\nu}^2 \,\, E_{e} \,\, p_{e} \times F(E_{\nu}) \, .
\end{eqnarray}
The neutrino energy goes from zero to to the maximal energy
\begin{equation}
E_{\nu}^{max} = (K + Q) \left( 1 - \frac{K+Q}{4 m_{p}+2K}\right) 
\approx K+Q \, ,
\end{equation}
where $K$ is the relative kinetic energy of the incoming protons, the
approximate equality disregards the recoil energy consistently with
Eq.~(\ref{spectrpp}), and where $F$ is the Fermi function
\begin{eqnarray}
F(E_{\nu}) 
            &=& \frac{ 2\pi\alpha \, E_{e} / p_{e} }
                      {\exp\left[ 2\pi\alpha \, E_{e} / p_{e} \right] - 1 } 
            \nonumber \\
&=&    \frac{
                 2\pi\alpha ( K+Q+m_{e}-E_{\nu} )
                      \left[(K+Q -E_{\nu})(K+Q+2m_{e}-E_{\nu})\right]^{-1/2}
                  }
             {\exp{\left[2\pi\alpha ( K+Q+m_{e}-E_{\nu} )
                     \left[(K+Q -E_{\nu})(K+Q+2m_{e}-E_{\nu})\right]^{-1/2}
                       \right] -1}
             }
 \, ,
\end{eqnarray}
which takes into account the Coulomb repulsion of positron and deuteron
in the final state.

The neutrino energy spectrum is obtained by convoluting this
result with the distribution for $K$: 
$S(E_{\nu})\sim \int dK P_{K}(E_{\nu}) f(K)$. In Fig.~\ref{figppsp},
we show the
energy distribution of the neutrinos from the $pp$ reaction for three cases:
(1) $T=0$ (dotted curve), (2) $T= 1.36$~keV with a Maxwellian distribution 
(solid curve) and
(3)  $T= 1.36$~keV with a $q=1.2$ Tsallis' distribution (dashed curve).

Note that relative large value of $|q-1|\approx 0.1 $ are necessary in
order that the effect on the spectrum of having a different distribution
be of the same order of magnitude of the standard thermal effect.
 
In principle, precise measurements of the neutrino spectrum give
information on the thermal distribution of the proton kinetic energies
$f(K)$. Let us estimate the necessary accuracy.
The central temperature of the Sun is about $kT_{c}=1.36$~keV and the
most effective energy $E_0$ (maximum of the Gamow peak) is about five times
as big for the $pp$ reaction (see $\tau$ in Table~\ref{table1});
therefore, thermal effects give a relative change of the maximal energy
about $5 \cdot 1.36 / 420 \approx 0.02$ and, in general, modify the shape
of the spectrum by a few percent. The center of mass motion gives also
contributions of order $kT$. Spectral measurements with accuracy better
than 1\% start to be sensitive to the particle thermal motion and
can measure the solar internal temperature. However, measurements that are
two orders of magnitude more accurate are necessary to measure the temperature
with a few percent error and/or discriminate between the
Maxwellian distribution and those slightly distorted distributions 
($|q-1|\approx 0.01$ or less) that we
shall consider in the next paragraph. Such kind of experiments are not
expected to be feasible in the near future.

\section{Solar neutrino experiments and the solar neutrino problem}
\label{secSNP}
\subsection{The solar neutrino problem}
The solar neutrino problem is one of the most interesting long standing puzzle
of the modern physics. The combined results from the solar neutrino
experiments (Homestake, GALLEX, SAGE, Kamiokande and SuperKamiokande)
cannot be reconciled with the predictions of the standard solar models (SSM).
Since SSMs have been very successful in predicting the stellar structure and
have given excellent descriptions of measurements as detailed
and accurate as the helioseismic ones, the neutrino experiments have
suggested that the minimal standard electroweak model should be extended
and include small neutrino masses and lepton flavor 
nonconservation~\cite{bahcall95b,castell97}. Neutrino oscillation theory
would have far-reaching consequences for both
particle physics and cosmology: therefore, it is of great importance to
ask whether the solar  neutrino problem can be solved, or at least
alleviated, in the framework of the conventional physics~\cite{dar97}. 

The problem can be more precisely appreciated with the help of 
Fig.~\ref{figsolnusp} and Table~\ref{table3}. Figure~\ref{figsolnusp}
show the energy spectrum of the main neutrino fluxes coming out of the
Sun. The same figure shows also the part of the spectrum that contributes
to each of the three present neutrino measures.
The energy dependence of each single flux depends practically
only on nuclear physics and it is independent of the SSM (see the
discussion about the $pp$ spectrum in the Sec.~\ref{secspectrum}).
However, the
integrated fluxes, which are reported in the first column of
Table~\ref{table3}, and, therefore, the relative weight of the different
components of the spectrum, depend on the SSM. 

One should compare these predictions with the experimental data (last
column of Table~\ref{table3}).

Only one experiment, the one performed
by Kamiokande and SuperKamiokande, (detection of the Cherencov light
emitted in water by electrons that are elastically scattered by solar
neutrinos, for the latest updated results see Ref.~\cite{suzuki98}) 
measures neutrinos from a single solar reaction
($^8$B~$\to 2\alpha + e^{+} + \nu_{e}$, the so-called boron neutrinos),
since its threshold (see Fig.~\ref{figsolnusp}) is too high to see any of
the other lower energy neutrinos. Therefore, its determination of
the boron flux is solar-model independent;  its present result corresponds
to less than one half of the predicted flux. This experiment has
measured also the energy spectrum of the boron neutrinos confirming the
expected shape apart for same discrepancy for energies greater than about
14~MeV, discrepancy that could be interpreted as evidence for
oscillations~\cite{bere98}.

The other two kinds of (radiochemical)
experiments~\cite{lande98,kirsten98,gavrin98}
have lower thresholds and, since
they do not have energy information, measure a combination of several
neutrino fluxes. Their interpretation in terms of individual fluxes
partially depends on the relative weight of the fluxes in
the solar model. However, they also show lower signals than predicted.
For a review on the SNP see Refs.~\cite{bahcall95b,castell97}.

\subsection{Standard attempts of astrophysical solutions}

The SSM flux predictions have ranges of variability, whose magnitudes
mostly depend on their importance in the energy production mechanism. 
The $pp$ flux is almost independent of the SSM (a few percent uncertainty),
the $^7$Be is fairly stable (about 10\% uncertainty), while the boron
flux, which is produced in a marginal chain of the energy production,
has a larger (more than 30\%) uncertainty. The ranges of predictions of
the SSM fluxes come mainly from the experimental/theoretical uncertainties
in the input parameters (nuclear cross sections, photon opacities
and initial contents).

The have been numerous attempts of reconciling solar models and neutrinos
experiments by allowing all theoretical inputs to vary within and also
considerably outside their uncertainties. Others have assumed arbitrary
changes of the internal structure of the models (scaling of the temperature
profile, fast element mixing, etc.).
This approach to the SNP is called the ``astrophysical solution''
(references could be found in Ref.~\cite{castell97}).

The most radical {\em astrophysical solution} consists in leaving the
neutrino fluxes as free parameters (no solar model) with
the only constraint that the produced energy should match the solar
luminosity. Nevertheless, there remains a discrepancy at about
the 3--4~$\sigma$ level (the central predictions of the SSM are more than
$10\sigma$'s from the experimental 
results)~\cite{hata94b,heeger96,bere96,castell97,bahcall98}.

\subsection{Non standard velocity distributions}
A different velocity distribution yields a different $\langle v\sigma\rangle$,
as discussed in Sec.~\ref{secmodirate}, and, therefore, a different solar
model and different neutrino fluxes.

In principle, the same variation of $\langle v\sigma\rangle$ that
is obtained by a different velocity distribution can also be
obtained by an appropriate change of the cross section, for instance by
means of the astrophysical factor $S$ in Eq.~(\ref{sigmas}). 
Therefore, allowing a modified velocity distribution does not span a
different set of neutrino fluxes than the arbitrary variation of all
the astrophysical factors. In particular, one cannot hope to achieve a better
agreement with the data than the one achieved by leaving the fluxes themselves
as free parameters.

In practice, however, arbitrary large variations of the astrophysical
factors contrast with our theoretical understanding of the nuclear
reactions and often with experimental measurements. Allowing
generous, but not arbitrary, ranges for the input parameters, implies a
much more restricted values of the possible neutrino fluxes.

On the contrary, small changes of the velocity distribution can give huge
changes of $\langle v\sigma\rangle$. Therefore, changing the
velocity distribution has several motivations. (1) When one is able
to calculated the velocity distribution from the microscopic physics, it
gives a predictions for $\langle v\sigma\rangle$. (2) Even when one can
only estimate the size of the deviation, it gives a physical mechanism
to justify large and correlated changes of $\langle v\sigma\rangle$.
(3) In any case, a given uncertainty in form of the distribution
yields calculable greater ranges of solar model predictions compared
to the ones of the SSM, which assumes the distribution function
known to high accuracy.

Changing $\langle v\sigma\rangle$ for the $i$th reaction will affect the whole
solar model and, in general, all fluxes will change. We use
Eq.~(\ref{asyratedel}) as a general parameterization of the rate change
for small deviations from the standard distribution. The effects of modifying
the rates on the fluxes can be estimated by using the power-law dependences
\begin{equation}
\label{dflux}
R_j\equiv\frac{\Phi_j}{\Phi_j^{(0)}}
  =\prod_i \left(
 \frac{\langle v \sigma_i\rangle_{\delta}}
      {\langle v \sigma_i\rangle_{0}}
                                 \right)^{\alpha_{ji}} 
     = e^{-\sum_i \delta_i\gamma_i \alpha_{ji} }  
\, ,
\end{equation}
for the fluxes $j=$ $^7$Be, $^8$B, $^{13}$N and $^{15}$O, and using
the solar luminosity constraint~\cite{castell97}
to determine the $pp$ flux,
$R_{pp} = 1+0.087\times(1-R_{\text{Be}})
                       +0.010\times(1-R_{\text{N}})
                       +0.009\times(1-R_{\text{O}})$, and keeping fixed
the ratio $\xi\equiv\Phi_{pep}/\Phi_{pp}=2.36\times 10^{-3}$.
The exponents $\alpha_{ij}=\partial\ln\Phi_j 
  / \partial\ln\langle v \sigma_i\rangle$ (see Table~\ref{table2}) have
been taken from Ref.~\cite{castell97}, where it is
also discussed why solar models depend only on the combination
$\langle v \sigma \rangle_{34} / \sqrt{\langle v \sigma \rangle_{33}}$
and why it is a good approximation to keep the ratio $\xi$ constant.

A direct microscopic calculation would determine $\delta$ that
could be different for every reaction ($\delta\to \delta_i$). The
energy distribution can be influenced by the specific properties of
the ion (charge and mass) and by the different conditions of the
environment in those parts of the Sun where each of the reactions
mostly takes place. However, such a direct calculation is not simple and
it does not exist for the solar interior. Therefore, for the purpose of
estimating
the effect of nonstandard distributions, we consider two simple models
and consider $\delta$('s) as free parameter(s). The first model assumes
the same deviation $\delta$ for all distributions, the second model
assumes that only the $p+{}^7$Be and ${}^3$He + ${}^4$He relative
energy distributions are nonstandard and introduces $\delta_{(17)}$
and $\delta_{(34)}$ to parameterize their deviations.

In the first case, one finds by substituting Eq.~(\ref{edg}) into  
Eq.~(\ref{dflux}) that
\begin{equation}
\frac{\Phi_j}{\Phi_j^{(0)}}=e^{-\delta \beta_j} \, ,
\end{equation}
where $\beta_j=\sum_i \alpha_{ji}\gamma_i$ are reported
in Table~\ref{table2}.
This dependence of the fluxes on $\delta$ is in good agreement with
 Clayton's numerical calculation~\cite{clayton75}.
Using the model of Ref.~\cite{bahcall95a} as reference model and the
experimental results up to the end of 1997 (see Table~\ref{table3}), we
obtain the best fit for $\delta=0.005$ with a $\chi^2=35$.

In the second case, we proceed similarly, but we use
$\delta_{(17)}$ for the reaction $p+ {}^7$Be and $\delta_{(34)}$  for the
reaction $^3$He + $^4$He:
the corresponding $\beta_j^{(17)}=\sum_{ i=17 }
\alpha_{ji}\gamma_i$ and $\beta_j^{(34)}=\sum_{ i=34 }
\alpha_{j,34}(\gamma_{34}-\gamma_{33})$ are also reported in
Table~\ref{table2}. As shown in Table~\ref{table3} the best fit is
obtained for $\delta_{(17)}=-0.018$ (negative $\delta$ corresponds to
an enhanced tail, $q>1$ in Tsallis' distribution)
and $\delta_{(34)}=0.030$ with a $\chi^2=20$.

As expected from the discussion in the first parts of this Section,
this result is a solution to the SNP, in the sense of providing a model
to fit the experimental results within one ({\em a few}) sigma.
However, we have shown that deviations from standard statistics corresponding
to values of $\delta$ of about 1\% can change the neutrino fluxes
of factors comparable to those that constitute the SNP.
Perhaps, the actual values of the neutrino fluxes coming out of the Sun
could result from the interplay of several mechanisms that are
disregarded in the standard picture~\cite{dar97}.

Moreover, it is clear that the uncertainties of the neutrino fluxes
are considerably underestimated by not considering the possibility of
non-extensive distributions.

\subsection{Helioseismic constraints}
Helioseismology provides very detailed and precise information on the solar
structure. The extremely precise measurements of a tremendous number of
frequencies give the possibility of extracting values of the sound speed
even near the solar core where the energy is generated.
In addition, several properties of the convective envelope are accurately
determined.

It is important to verify that nonstandard distributions do not contrast
with such data. This study has been done for the velocity distribution
of proton~\cite{degli98b}: if $\delta$ is within with following limits
\begin{equation}
\label{rangedelta3s}
  -4.9 \times 10^{-3} < \delta_{pp} < 2.3 \times 10^{-3} \, 
\end{equation}
there is no incompatibility between helioseismic data and
nonstandard statistics.

As we have seen, even such small values of $\delta$ have very important
consequences for the neutrino fluxes; in addition, this limit does not
automatically apply to the distribution of other ions.

\section{Conclusion}
\label{secconclu}
The solar core, where its energy is produced, is a weakly nonideal
plasma.

Many approaches are possible to such systems. The standard
Debye-H\"uckel theory is very successful, but it is not sufficient
to give an accurate description when one needs to calculate the
energy (velocity) distribution function to a high accuracy.

We have tried several new approaches to this complex problem.
In particular, we have considered corrections to the Fokker-Plank
equations, the known stochastic distribution of electric microfields,
and memory effects arising from the interaction of individual and collective
variables.

All our attempts indicate that the velocity distribution of ions in the
plasma could be different from the Maxwellian one, and should be
well described by a Tsallis' distribution slightly nonextensive
($|q-1|\sim 0.01$).

Such small deviations from the standard statistics produce
effects on the energy dependence of the neutrino spectra or on the 
helioseismic observables that are not in contrast with present data.

However, even such small deviations of the energy distribution produce
dramatic effects on those nuclear rates whose main contributions come from
the high-energy tail of the distribution, as it is best exemplified by
Fig.~\ref{figgamow}.

This theoretical possibility enlarges the range of predictions for the
solar neutrino fluxes and, while it is not sufficient to solve the
solar neutrino problem, can make it somewhat less dramatic.

\begin{table}
\caption[taa]{
 Most effective energies for thermonuclear reactions and exponents
 $\gamma$ that characterize the change of the
 thermal average $\langle v \sigma \rangle$ to the leading order in
 $\delta$, when the energy distribution changes by a factor
 $\exp\{ -\delta (E/kT)^2 \}$:
 $\langle v \sigma \rangle_{\delta} =
  \langle v \sigma \rangle_{0} \exp\{ -\delta\gamma \}$.
\label{table1}
             }
\begin{tabular}{rcdd}
\multicolumn{2}{c}{reaction} &
         \multicolumn{1}{c}{$\tau=E_0/kT$} &
           \multicolumn{1}{c}{ $\gamma = \tau^2$} \\
\tableline
 $\langle v \sigma \rangle_{11}$:  &
        $p + p\to {^2\text{H}} + e^+ + \nu $  &
                4.8  &  23. \\
 $\langle v \sigma \rangle_{17}$:  &
        $p + {^7\text{Be}} \to {^8\text{B}} + \gamma $  &
               13.8  &  190. \\

 $\langle v \sigma \rangle_{33}$:  &
        $^3\text{He}+  {^3\text{He}}\to \alpha + 2p $  &
               16.8  &  281. \\
 $\langle v \sigma \rangle_{34}$:  &
        $^3\text{He}+{} ^4\text{He}\to {^7\text{Be}} + \gamma $  &
               17.4  &  303. \\
$\langle v \sigma \rangle_{1,14}$:  &
       $p + {^{14}\text{N}} \to {^{15}\text{O}} + \gamma $  &
               20.2  &  407. \\
\end{tabular}
\end{table}
\begin{table}
\caption[tbb]{
The first four rows show $\alpha_{ij}=\partial\ln\Phi_{j}/
\partial\ln \langle v \sigma \rangle_{i}$, the
logarithmic partial derivative of neutrino fluxes with respect to the
parameter shown at the left of the row. These numbers are discussed in
Ref.~\cite{castell97}. The last three rows show $\beta_j$,
$\beta_j^{\text{Be}}$ and $\beta_j^{\text{He}}$, the logarithmic partial
derivative of the fluxes
with respect to the parameters $\delta$'s; as discussed in the text, they
are linear combinations of the $\alpha$'s weighted by the factors
$\gamma$ of Table~\ref{table1}.
\label{table2}
             }
\begin{tabular}{lccc}
 & $^7$Be & $^8$B & CNO \\
\tableline
$\langle v \sigma \rangle_{11}$ &
         -1.0   &   -2.7   &   -2.7   \\
$\langle v \sigma \rangle_{34} / \sqrt{\langle v \sigma \rangle_{33}}$ &
         +0.86   &   +0.92   &   -0.04   \\
$\langle v \sigma \rangle_{17}$ &
         0   &   1   &   0   \\
$\langle v \sigma \rangle_{1,14}$ &
         0   &   0   &   1   \\
\tableline
$\beta_j$ &
         117   &  277   &   338.5   \\
\tableline
$\beta^{\text{Be}}_j$ &
          0   &  190   &   0 \\
$\beta^{\text{He}}_j$ &
         140   &  150    &  -6.5   \\
\end{tabular}
\end{table}
\begin{table}
\caption[tcc]{
The first three columns show the predicted fluxes, and the
predicted gallium and chlorine signals in the SSM~\cite{bahcall95a} and
in the two models with nonstandard distribution described in the text.
The last column shows the present experimental results. For the three
models is also given the $\chi^2$ resulting by the comparison with the
experimental data.
\label{table3}
             }
\begin{tabular}{ccccc}
&  \multicolumn{3}{c}{Models}   &   \\
\cline{2-4}
&  SSM          & case I       & case II &  Experiment           \\
& ($\delta=0$)  & ($\delta=0.005$)
                & ($\delta_{\text{Be}}=-0.018$,
$\delta_{\text{He}}=0.030$) &           \\
\tableline
$[10^9 \text{cm}^{-2} \text{s}^{-1}]$ &
             &     &      &  \\
$\Phi_{pp}$ &
        59.1  & 62.2  &  63.7  &  \\
$\Phi_{^7\text{Be}}$ &
         5.15  &  2.87  &  0.08  &  \\
$\Phi_{^{13}\text{N}}$ &
         0.62  & 0.11  &  0.75  &  \\
$\Phi_{^{15}\text{O}}$ &
         0.55  & 0.10  &  0.67  &  \\
$[10^6 \text{cm}^{-2} \text{s}^{-1}]$ &
             &     &      &  \\
$\Phi_{^8\text{B}}$ & 6.62 & 1.65  &  2.25  &
             $2.55\pm 0.21$~\tablenote{Weighted average of
         $2.80\pm 0.38$~\protect\cite{fukuda96}
     and $2.44\pm 0.26$~\protect\cite{kataup97}}\\
\tableline
[SNU] &
             &     &      &  \\
gallium & 137.0  & 100  &  97  &
   $75\pm 5$~\tablenote{Weighted average of
   $76\pm 8$~\protect\cite{gataup97} and
 $ 72\pm 13$~\protect\cite{gavrin96} }\\
chlorine &
         9.3  & 2.84  &  3.34  &
$2.54\pm 0.20$~\tablenote{Ref.~\protect\cite{lande96}} \\
\tableline
$\chi^2$ &
         74  & 35  &  20  &     \\
\end{tabular}
\end{table}
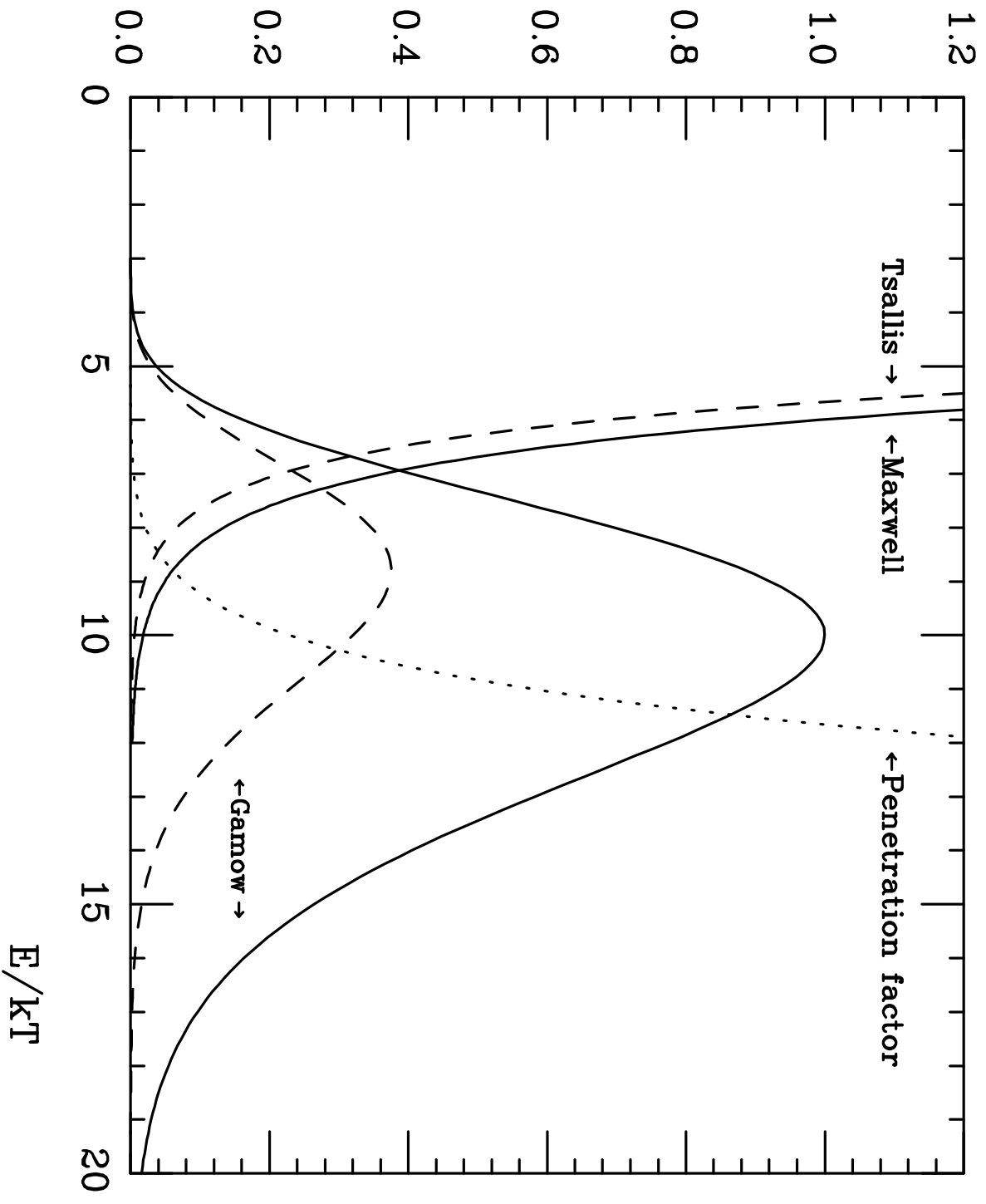
\begin{figure}
\caption[gamo]{
\label{figgamow}
The Gamow peak for the Maxwellian and the Tsallis' distribution.
The exponentially decreasing solid (dashed) curve labeled Maxwell (Tsallis)
shows the Maxwellian (Tsallis with $q=0.98$) energy distribution. Both the
energy distributions have been normalized to a huge 400 at
$E/k T=0$, to emphasize their tiny difference (the part of the distributions
shown contains only about 0.3\% of the total area).
The rapidly increasing dotted curve shows the behavior of the penetration
factor $\exp{-\sqrt{E_G/E}}$ with $E_G = 4000 kT$ corresponding to
$\tau=E_0/kT = 10$; the normalization is arbitrary. The solid (dashed)
peak shows the product of Maxwellian (Tsallis) distribution function
times the penetration factor.
  }
\end{figure}
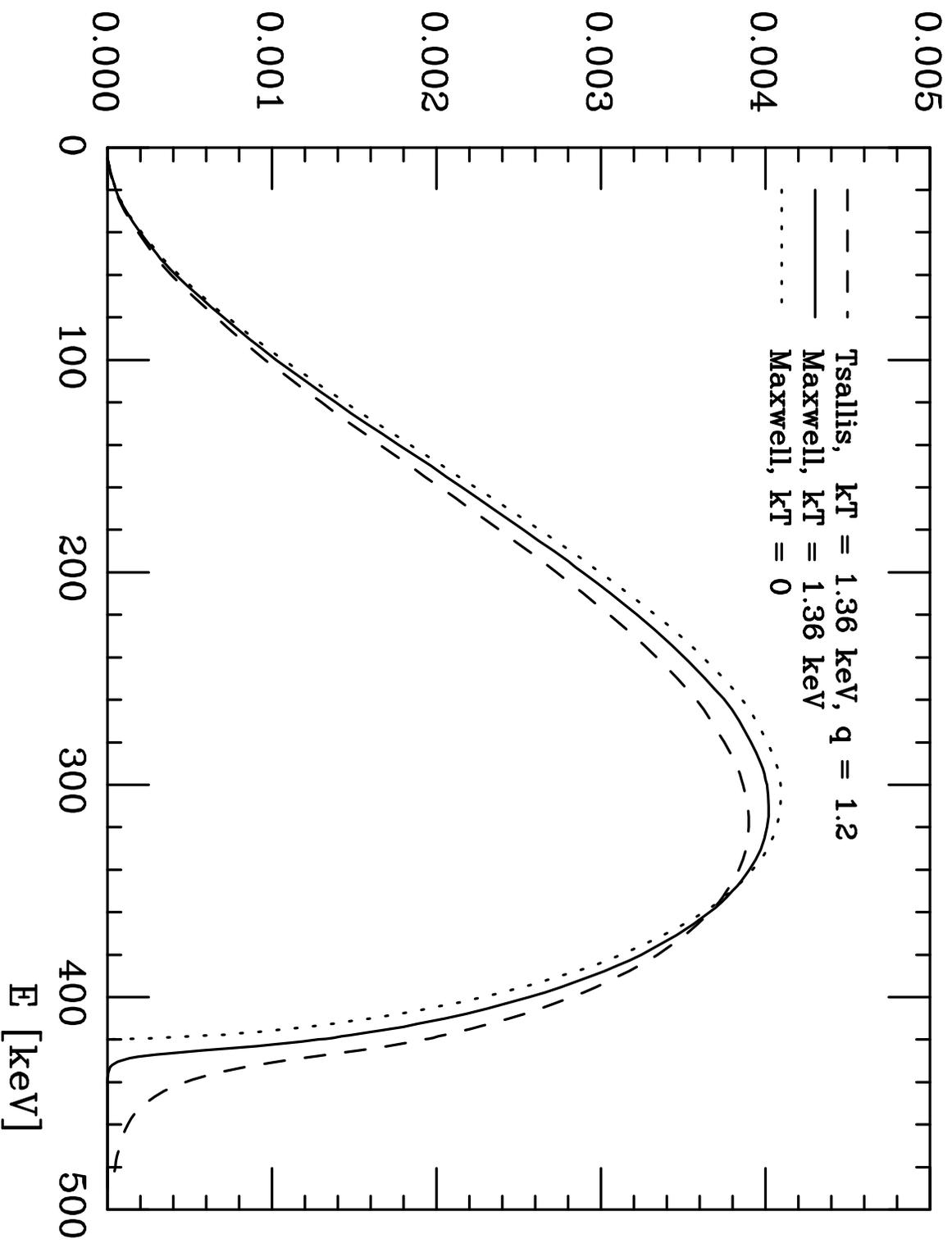
\begin{figure}
\caption[pp]{
The spectrum of solar neutrinos produced by the $pp$ reaction. The solid line
shows the spectrum at $kT= 1.36$~keV (about the solar central
temperature) with the Maxwellian distribution for the proton relative
energy, the dashed line the spectrum with the Tsallis' distribution at
the same temperature. For comparison the dotted line show the spectrum
with a Maxwellian distribution at $T=0$.
 \label{figppsp}
  }
\end{figure}
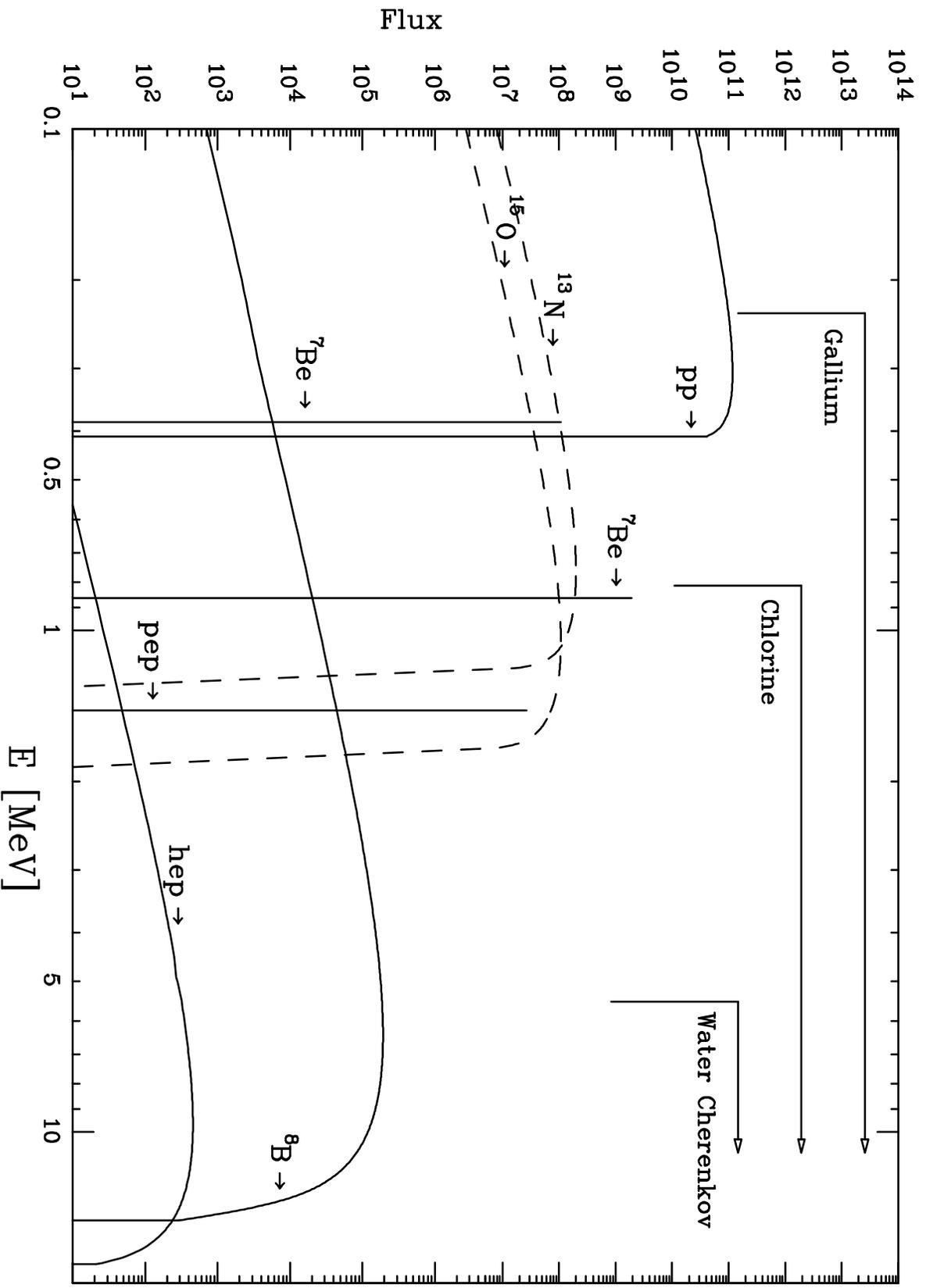
\begin{figure}
\caption[sunu]{
The solar neutrino spectrum. For continuous sources, the differential flux
is in cm$^{-2}$s$^{-1}$~MeV$^{-1}$. For the lines, the total flux is in
cm$^{-2}$s$^{-1}$. There are also indicated the thresholds of the three
kinds of present experiments. The radiochemical experiment at
Homestake~\cite{lande98,lande96} that uses using the reaction
$\nu_{e} + {}^{37}$Cl $\to {}^{37}$Ar (Chlorine) has the threshold
0.814~MeV. The radiochemical experiments
Gallex~\cite{kirsten98,gataup97} and SAGE~\cite{gavrin98,gavrin96}
that use the reaction $\nu_{e} + {}^{71}$Ga $\to {}^{71}$Ge (Gallium) have
the threshold 0.2332~MeV. The thresholds of the Kamiokande and Superkamiokande
experiments~\cite{suzuki98,fukuda96,kataup97},
which detect the Cherencov light emitted by the scattered
electrons in water (Water Cherenkov) are chosen between 5.5 and 8~MeV.
 \label{figsolnusp}
  }
\end{figure}
\end{document}